# Understanding: reframing automation and assurance

Robin E Bloomfield
*City St Georges, University of London*
London, UK
r.e.bloomfield@citystgeorges.ac.uk

*Abstract*—Safety and assurance cases risk becoming detached from the understanding needed for responsible engineering and governance decisions. More broadly, the production and evaluation of critical socio-technical systems increasingly face an understanding challenge: pressures for increased tempo, reduced scrutiny, software complexity, and growing use of AI-generated artefacts may produce outputs that appear coherent without supporting genuine human comprehension. We argue that understanding should become an explicit, assessable, and defensible component of decision making: what developers, assessors, and decision makers grasp about system behavior, evidence, assumptions, risks, and residual uncertainty. Drawing on Catherine Elgin's epistemology of understanding, we outline a conceptual foundation and then use Assurance 2.0 as an engineering route to operationalize using structured argumentation, evidence, confidence, defeaters, and theory-based automation. This leads to two linked artefacts: an Understanding Basis, which justifies why available understanding is sufficient for a decision, and a Personal Understanding Statement, through which participants make their grasp explicit and challengeable. We also identify risks that automation may improve artefact production while weakening understanding, and we propose initial directions for evaluating both efficacy and epistemic impact.

*Keywords— Understanding, assurance cases, safety cases, automation, formal arguments, AI assurance, Assurance 2, epistemic justification, certification.*

## I. Introduction

This paper identifies and responds to a contemporary challenge of understanding in the production and evaluation of critical socio-technical systems. One driver is the need for assurance-related understanding to support responsible and accountable decisions to deploy and operate systems. Another is the need for software understanding sufficient to verify system functionality, safety, and security.

Safety and assurance cases, and other approaches to evaluation, risk becoming detached from the understanding needed for responsible engineering and deployment decisions: they are in danger of losing, or already have lost, the plot [1]. Engineering decision making increasingly takes place in contexts shaped by socio-technical complexity, rapid development tempos, organizational change, and growing use of automation and AI [2]. In such settings, existing assurance practices may come to optimise the production of documentation rather than the support of justified judgement. The concern is broader than assurance cases.

Software understanding presents a parallel challenge. A joint initiative from CISA, DARPA, OUSD(R&E), and NSA identified a "software understanding gap": investment in software production has outstripped investment in understanding for decades, leaving mission owners and operators unable to verify the functionality, safety, and security of systems they depend on [3]. The focus [3] is secure by design and on mathematically proven software. Yet even if software is proved correct – which would itself deliver a massive change in our confidence – we need to know the applicability of models and their validity, to understand how doubt in formulating these proven properties is dealt with and the impact of doubts in the tool chain. In addition, we need to consider how this information is used to make decisions and how it can be communicated out of the specialist areas that generate the evidence.

The concern is not only whether automation can generate assurance material efficiently. It is whether automation preserves, improves, or undermines the understanding that developers, assessors, and decision makers need in order to act responsibly. Automation may increase the speed, quantity, and apparent consistency of assurance artefacts while reducing the degree to which any human actually understands the system, the assumptions, the evidence base, or the residual risks. Recent work in AI-assisted programming has introduced the notion of epistemic debt: the accumulation of functional capability without corresponding growth in the mental models needed for correction and maintenance [4]. Although this evidence comes from programming rather than the wide range of disciplines involved in developing and assuring critical systems, it highlights a risk that automation may support immediate output while weakening later understanding and intervention. Furthermore, our ability to recognize this debt is hampered by the illusion of explanatory depth, a cognitive bias where individuals believe they possess a detailed, coherent understanding of a complex system, when in reality their grasp is shallow until challenged to provide a step-by-step explanation [5]. This matters because reviewers may mistake familiarity, coherence, or fluent tool output for genuine grasp, only discovering the shallowness of their understanding when required to explain assumptions, mechanisms, evidential relevance, or conditions for revision.

For decisions about whether to deploy a system, accept residual risk, or approve a change, two questions matter. The first concerns efficacy: how far automation improves the creation, maintenance, and review of the artefacts that support a decision. The second concerns epistemic quality: how far those artefacts support genuine understanding of hazards, threats, assumptions, uncertainties, and trade-offs. Addressing these questions can pull in opposite directions: a process may become more efficient while the resulting understanding becomes thinner, more fragile, or more deferential to apparently coherent outputs. Generalizing a Fundamental Principle from the UK nuclear Safety Assessment Principles (SAP) [6], we assert that a core requirement is that human decision makers should *understand* the relevant risks well enough to make valid decisions and accept responsibility and accountability. The mechanism to demonstrate this is, in fact, a safety case [7]. The SAPs and associated guidance discuss the role of safety cases as a mechanism for showing understanding. So this goes right back to the origins of what a case is for: demonstrating and communicating understanding.

Yet understanding is usually treated as implicit, personal, and indirectly evaluated. Conventional assurance approaches focus on claims, arguments, evidence, and confidence while development processes focus on verification and validation artifacts. Neither make explicit what relevant actors understand, how that understanding is formed, how it is challenged, or how it changes when new evidence or defeaters arise. This matters because collecting evidence is not the same as achieving understanding. Nor is internal coherence sufficient by itself. In high-consequence settings, a persuasive account that is poorly understood can be as problematic as an unsupported one.

This becomes more problematic as decision making is increasingly mediated by tools that can generate, transform, and summarize assurance content at scale. The understanding gap is further compounded by generational loss of expertise, outsourcing and supply-chain dependence as well as the pernicious ubiquity of the illusion of explanatory depth.

This paper argues that understanding should become an explicit, assessable, and defensible component of engineering decision making, particularly for critical and high-consequence systems. Understanding is not a simple matter of collecting more facts: it has philosophical, psychological, pedagogic, and practical dimensions. The core of our approach is to draw on Catherine Elgin's epistemology of understanding [8] to develop a conceptual account that is rich enough to guide practice while still being operationally useful. To make that practical, we introduce two linked artefacts: an Understanding Basis, a structured justification of why the available understanding is sufficient for the decision at hand and how that judgement remains open to challenge and revision and a Personal Understanding Statement, through which a participant in a role makes their understanding explicit in a disciplined way.

The paper makes three contributions. First, it diagnoses a failure mode in contemporary assurance practice: a drift towards document production that can be intensified by automation. The prevailing framing treats automation as a production problem: can we generate assurance artefacts faster, more consistently, and at lower cost? We propose a different framing: automation must also be evaluated by whether it enhances or degrades the human understanding that makes those artefacts meaningful. Second, it proposes understanding as an explicit object of assurance, using Elgin's work to clarify what such understanding involves. Third, it relates these ideas to Assurance 2.0 [9], suggesting how structured argumentation, evidence, confidence, defeaters, and automation might be redesigned so that they support accountable human understanding rather than merely the appearance of assurance.

## II. PHILOSOPHICAL FOUNDATIONS[1]

We all know intuitively what understanding is, but how do we operationalize it so we can evaluate and communicate what we understand? We turn to the work of the contemporary philosopher Catherine Elgin [8] to provide the conceptual basis. Elgin explores a general notion of understanding that is widely applicable to science, thought experiments, mathematics, novels, dance, and fine art, and in each case different aspects of understanding are emphasized or prioritized.

Elgin uses two metaphors: grasping and fabric. "The term 'understanding' as a success term for having an epistemically suitable grasp of or take on a topic". The difficulty, as she notes, is to spell out in nonmetaphorical terms what grasping is. An "important element of grasping is knowing how to exploit the information or insight one's understanding provides. Someone who understands a proposition knows how to wield it to further her cognitive (and perhaps practical) ends. Someone who understands a topic knows how to use the epistemic resources her take on that topic affords."

The grasping metaphor also suggests that understanding is non-linear. We speak of a preliminary understanding as "getting the gist of something"; there comes a time when we grasp it. We can go beyond grasping to more expert use. Grasping refers to an inner state, feeling of understanding and that can be demonstrated and communicated by how we explain, empathize, predict. In this paper we recognize this inner state and look for externalities that allow us to develop, communicate and challenge that understanding.

She talks about the "The Fabric of Understanding" and for her an account is more than a thin "golden thread" from evidence to conclusion, it is a tightly interwoven tapestry of mutually supportive commitments. To quote Elgin "At a first approximation, an understanding is an epistemic commitment to a comprehensive, systematically linked body of information that is grounded in fact, is duly responsive to reasons or evidence, and enables nontrivial inference, argument, and perhaps action regarding the topic the information pertains to". In our context, this resonates with a definition of an assurance case as "a documented body of evidence that provides a convincing and valid argument that a system is adequately dependable for a given application in a given environment" [10].

A key insight of Elgin is that understanding is based not on literal truth but on things that are "true enough". . Scientific and engineering understanding often relies on idealizations, abstractions, simplifications, and models that are not strictly true. Elgin's notion of **felicitous falsehoods** captures this directly. Such representations can be valuable, not despite their departures from truth but because selective distortion can reveal what matters. In complex domains, a fully detailed account may obscure rather than illuminate. A model may omit, smooth, aggregate, or stylise in order to make salient the structure of a phenomenon, the interaction of variables, or the limits of a design. In this sense, understanding depends on being *true enough* for the purpose at hand. For assurance, this is important because many of the representations we rely on—hazard analyses, architecture views, operational scenarios, confidence arguments, and risk models—are selective constructions. Their value lies not in exhaustive factual completeness but in whether they support sound understanding and defensible judgement. Her notion of "**felicitous falsehoods**" is much stronger than just acknowledging the limitations of our models or knowledge. It is a fundamental property that in order to be useful, any claim or argument has to be selective in what it is referring to: they are ineliminable valuable components of understanding.

---

[1] An engineering-oriented glossary is provided in Section VI.

She then looks to **coherence -** internal consistency and mutual support among the elements of an account – to provide justification for the claims and then looks at how this one account relates to other related accounts. This results in a **reflective equilibrium in** which each element of the network is reasonable in light of the others, and the network as a whole is as reasonable as any available alternative in light of our relevant previous commitments. Even if some components would be doubtful in isolation, collectively they constitute an interwoven tapestry of commitments that we can on reflection endorse. Reflective equilibrium is both a method and a state. It is a method because inquiry proceeds through revision, comparison, challenge, and repair. It is a state because at any point we may arrive at a position that is sufficiently integrated and defensible for present purposes. In assurance terms, this resonates with the iterative refinement of claims, evidence, assumptions, confidence judgements, and rebuttals until the overall case is one that responsible participants can endorse.

The method of reflective equilibrium is dialectical and its results are provisional. Elgin argues that fallibilism should be, woven into the fabric of understanding. The possibility of error does not undermine understanding; rather, it is built into it. An account can be well supported, coherent, and useful while still remaining revisable. Indeed, mechanisms for identifying error, resolving conflict, and incorporating new evidence strengthen understanding rather than weaken it. For engineering assurance this is crucial. Assurance should not be framed as if it culminates in certainty. It should instead make visible where confidence is strong, where it is conditional, what assumptions carry weight, and how the account could change in the light of new evidence, anomalies, or challenge. The move from a focus on knowledge to a focus on understanding pays dividends as understanding provides both a way of evaluating the significance of error but also an ability to learn from it.

Elgin discusses at some length the difficulty in achieving coherence and reflective equilibrium for accounts that are wrong: to do this the confabulator has to ignore many aspects that they might normally consider. "But **coherence through confabulation** is not so easy to achieve as philosophers imagine…Confabulators achieve coherence among their first-order beliefs by sacrificing coherence between first- and second-order beliefs." In her metaphor, the account becomes threadbare. Nevertheless, conspiracies and other fictions can be consistent and convincing despite being fictitious.

Coherence alone, however, is not enough, we need to **tether the account to reality** through evidence and fact. One mechanism that tethers understanding to the world is **exemplification** – "the connection between a sample or example and whatever it is a sample or example of". A swatch of fabric in a catalog, for instance, is an exemplar of a particular texture and color. It both *is* that texture and color (instantiation) and *stands for* that texture and color (reference). In scientific and engineering work, test results, simulations, incidents, prototypes, operating experience, and worked examples can play a similar role when they genuinely instantiate the features for which they are taken as evidence. Exemplification helps explain how an account connects to the world rather than merely hanging together internally and resonates with the use of evidence assembly and evidential claims [11].

Elgin argues that "**accounts must be assessed holistically**. Rather than asking whether each component sentence expresses a fact, we should ask whether the account as a whole is in reflective equilibrium." She notes that collecting knowledge and facts is not understanding. Sometimes reducing the number of facts to make something that is less true but more insightful increases understanding. What is needed is a holistic account – she points out, in assurance case terms, that just amassing conjunctions of claims undermines confidence in the overall judgement. What is needed are fewer facts but theories that explain how claims give rise to justification.

The two concepts of coherence and exemplification can drive the process of inquiry. Coherence, achieved through reflective equilibrium, is the internal, structural property of an epistemic account. In dealing with complex systems we would differentiate between coherence external to our justification and internal. Exemplification tethers the account to the world through the instantiation of properties in its exemplars. An account that is merely coherent without being tethered by exemplars is a delusion. A collection of exemplars (e.g., raw data) that has not been organized into a coherent account provides no understanding.

She notes that the process of building understanding is a continuous cycle: we seek out exemplars from the world to ground our account, but we interpret what those exemplars mean and which of their many properties they exemplify through the lens of our currently coherent account. When a new exemplar—a new piece of evidence—resists incorporation, it disrupts the coherence of the account. This forces a revision of the account's commitments until a new, more comprehensive reflective equilibrium is achieved that can accommodate the new exemplar. This resonates with safety practice.

Elgin also discusses the adjudication, need for integrity, ethics and process and the dialectical aspects of argumentation all of which are part of the use of the assurance case process, at least in the engineering of critical systems.

For our purposes, five features of Elgin's account are important. First, understanding is a matter of grasp: being able to use an account to reason, explain, and act, not merely to recite it. Second, understanding has the character of a fabric: an interwoven structure of mutually supporting commitments rather than a linear chain. Third, felicitous falsehoods show why models and idealizations can advance understanding when they are appropriately selective. Fourth, tethering through exemplification explains why coherent reasoning must still be grounded in evidence that genuinely bears on the matter at hand. Fifth, reflective equilibrium makes understanding iterative and revisable, while fallibilism ensures that openness to error is treated as a resource for inquiry rather than a threat to it.

This philosophical framing is especially relevant where (a) complex engineering models use deliberate idealizations, (b) reasoning spans multiple disciplines, and (c) decisions must be revisable in light of new evidence.

A potential danger of the emphasis on coherence is a reluctance to admit genuinely new ideas, since the interwoven structure of reflective equilibrium has a conservative tendency toward stability. How then does understanding accommodate radical novelty — new threats, new science, new failure modes? Arendt's distinction between *preliminary* and *true* understanding [12] was precisely about this. She warns that when the world "breaks," our first impulse is to *force* coherence — to fit the unprecedented into old frameworks. But genuine understanding begins only when we resist equilibrium — when we acknowledge dissonance, rupture, and meaninglessness before trying to make sense. Disasters occur when the unprecedented is dismissed as impossible because it does not fit existing categories of thought: disasters occur because of failures of imagination. This is particularly germane today where we have unprecedented changes in technology and institutions.

We therefore have begun to integrate Elgin's epistemic notion of *reflective equilibrium* with Arendt's existential account of *understanding after rupture*. We use Arendt to highlight rupture: moments when established categories fail and forced coherence becomes misleading. This is particularly important in the context of rapid technological and institutional change, so we propose that understanding in engineering must support not only coherence and revision, but also explicitly recognize rupture and subsequent re-evaluation.

III. DEVELOPING AN OPERATIONAL FRAMEWORK

Elgin provides a conceptual basis for making claims about understanding. We want to develop a methodology based on these foundations. This is work in progress, and in this short paper we summarise the directions of the research in concept mapping, roles and artefacts, and automation experimentation.

We acknowledge that achieving understanding in engineering requires more than philosophical grounding. A holistic methodology would benefit from integrating philosophical, psychological, pedagogic, organizational, and modelling perspectives. Several directions are especially important: tackling distributed decision making, where an Understanding Basis would need to represent distributed understanding across roles and organizations; recognizing the centrality of models as epistemic tools that shape what is understood [13, 14, 15]; incorporating insights from other philosophers, including Peirce's triad of abduction, deduction and induction [16] as well as the role of imagination and intuition in framing alternatives; drawing on educational perspectives such as Wiggins & McTighe's "six facets of understanding" [17]; mitigating psychological pitfalls including groupthink, misplaced coherence, bias, and particularly the illusion of explanatory depth; exploring narratology, since an assurance case is not just a logical assembly of claims and evidence but a story, and work in security has shown that narratological approaches can make cases more intelligible and better suited to promoting understanding; and finally, applying the approach to itself, using an Understanding Basis to evaluate the methodology proposed in this paper.

*A. Concept mapping*

The concepts of Assurance 2.0 provide an engineering approach to Elgin's philosophy, moving from philosophy to practical engineering. We have mapped Elgin's epistemology onto features of Assurance 2.0 and vice versa, in summary, showing that:

- **Epistemic commitments** are expressed through structured claims, assumptions, theories, and justification.
- **Felicitous falsehoods** appear in model assumptions and decomposition strategies, which can be justified through side-claims and confidence assessments.
- **Coherence** is supported by structured argument blocks, consistency checks, and explicit handling of assumptions and evidence.
- **Tethering** is achieved with evidential claims and confirmation theory, which interpret evidence rather than merely list it.
- **Reflective equilibrium** and fallibilism is operationalized through defeater analysis, uncertainty management, dialectical challenge, and explicit documentation of reasoning.
- **Rupture and re-evaluation** may be supported by defeaters specifically designed to resist forced coherence together with explicit claims about confidence and expected future behavior.

While many features of Assurance 2.0 implicitly support understanding, there are areas where further emphasis or development would be valuable—for example, representing felicitous falsehoods more explicitly, capturing external coherence and novelty, addressing tension between reflective equilibrium and rupture and providing mechanisms for demonstrating "grasp".

*B. Roles and artifacts*

In practice, there may be many roles for which explicit understanding is important to a decision. Here we use two stylised roles to keep the discussion simple: a domain expert, who develops and maintains the justification, and a decision maker, who must decide, on the basis of both their own expertise and the account provided by others, whether the available understanding is sufficient for the decision at hand.

We propose two linked artefacts to embed understanding into the engineering lifecycle: an Understanding Basis and a Personal Understanding Statement. The Understanding Basis is a structured Assurance-2.0-style justification addressing why the available understanding is sufficient for a specific decision. The Personal Understanding Statement is a reflexive artefact through which a participant makes their grasp explicit, challengeable, and revisable.

There is a design question about how directly Elgin's concepts should appear in the claims themselves, and how far they should instead be translated into more familiar engineering and socio-technical terms. One possible approach would be to use Elgin-inspired ideas as higher-level objectives or principles, building on the general direction of principles-based assurance.

The Understanding Basis would operationalize the concepts outlined in the preceding concept mapping. In a decision-specific form structured in Assurance 2.0, it would make explicit how the justification is tethered to evidence, what epistemic commitments it relies on, how idealizations and felicitous falsehoods are being used, how coherence is established and challenged, and how confidence, defeaters, and the possibility of rupture are handled. In this sense, the concept mapping provides the framework, while the Understanding Basis is the concrete instantiation of that framework for a particular decision, system, and context.

In addition, the Basis should incorporate designed friction: deliberate pauses and prompts that require reviewers to interrogate complex reasoning rather than passively accept automated outputs. The aim is to mitigate shallow uptake, theory opacity, and the accumulation of epistemic debt. The Understanding Basis complements, but does not replace, system-focused assurance cases, but their relationship needs further work.

The Personal Understanding Statement extends the Assurance 2.0 "sentencing statement" idea by enabling decision makers and developers to articulate, with evidence, what they understand and why that understanding is sufficient. This includes understanding of system behaviour and context, of the evidence, models, and chains of reasoning involved, of both positive and negative evidence, of the doubts that remain and how they are handled, and of how judgements are communicated and revised.

Developing such a statement involves active analysis of the Understanding Basis. This includes examining the sensitivity and role of inductive and deductive reasoning, the participant's confidence in different parts of the Basis and in the overall judgement, whether felicitous falsehoods are adequately defined and justified, and how coherence and reflective equilibrium might be challenged, strengthened, or disrupted. It also includes how the assessor judges the adequacy, limitations, and vulnerability of the Basis to rupture.

In addition, in the spirit of coherence, the review should also interrogate the assessor's overall sense of the adequacy of the Basis, including any felt unease, tension, or lack of fit that may signal unresolved incoherence or unexamined assumptions.

There is also a methodological challenge in how review should be supported and tested. For example, reviewers might be presented with plausible but logically flawed automated outputs in order to expose theory opacity. They might be required to translate formal structures into a coherent narrative, to test whether the account is genuinely understood rather than merely followed. It may also be useful to assess whether a reviewer can still defend the Understanding Basis when automated assistance is removed, thereby revealing hidden epistemic debt.

*C. Automation evaluation – efficacy and understanding*

Of particular relevance to FACCT 2026 is the need to evaluate the impact on understanding of automation and formal reasoning. As part of the DARPA Clarissa project [9], we used a logic based language and associated tools to develop a Synthesis Assistant to automatically synthesis CAE-based cases, add identified evidence and relevant defeaters. It is based on theories as ontologies and logic-based transformations. The underlying motivations for synthesis automation include: (i) reduce errors in case construction, (ii) increase tempo of case production (and reduce cost), respond to DevSecOps and compile to combat challenges, (iii) remove unnecessary variability in cases, so easier to review, and (iv) discover design options, show trade-offs, support assurance as part of development work-flow.

The claims we make about the approach are that it shifts the review effort towards understanding theories, assessing their relevance and validity and the trustworthiness of the tools.

We also hypothesize that there is complexity reduction in the review effort because of the focus on theories ("the rest is knitting"). We also surmise that automated checks for unused evidence and components are beneficial and that the ability to generate all cases with respect to a constraint allows us to explore the possible solution space and alternatives.

This is summarized in the hypothesis tree below. Defining the tree addresses some of the Nasa critique [19] of the lack of rationale for using LLMs in assurance cases.

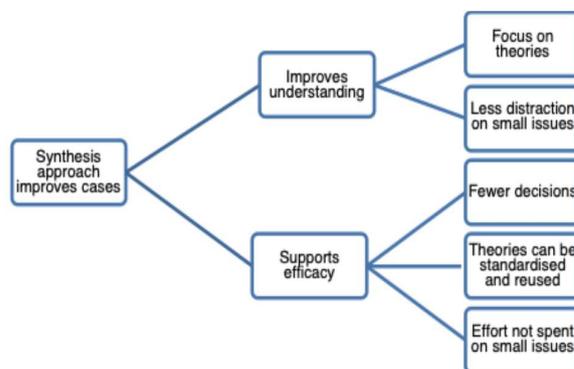

*Figure 1 Hypothesis tree for automation benefits – case study*

We have used the hypothesis tree to brainstorm a range of defeaters for the automation. The key defeaters identified in this analysis include: (1) LLM unreliability — generated case structures may be plausible but incorrect, with errors that are difficult for reviewers to detect; (2) theory scarcity — if reusable formal theories do not yet exist for a domain, synthesis provides no efficiency gains; (3) theory opacity — the understanding required to validate a theory may be no less demanding than reviewing the case itself; (4) review dynamics — the social and deliberative aspects of case review (challenge, discussion, unexpected insight) may be degraded when reviewers encounter a pre-formed automated output rather than building understanding through engagement with the argument structure; and (5) tethering loss — automated generation risks producing artefacts that are internally coherent but disconnected from the operational realities that give them meaning. This analysis has a direct implication for formal argument tools of the kind central to FACCT: such tools should be evaluated not only for whether they produce correct and complete artefacts, but for whether they support or supplant the epistemic grasp of the human reviewer. The prevailing framing treats automation as a production problem — can we generate assurance artefacts faster, more consistently, and at lower cost? We propose a complementary framing: does automation enhance or degrade the understanding that makes those artefacts meaningful? Artefact production and genuine comprehension are not the same thing and can pull in opposite directions. A formal argument tool that accelerates production while eroding the reviewer's ability to challenge, interpret, and take responsibility for the argument might be a net loss, even if the artefact is formally correct.

## IV. NEXT STEPS AND IMPLICATIONS

The work reported in this paper is preliminary and work in progress. As noted earlier, achieving understanding in engineering requires more than philosophical grounding. A fuller methodology will need to integrate philosophical, psychological, pedagogic, organizational, and modelling insights.

The next stage of the work is to elaborate the present framework further and to test and refine it be defining use cases and trials. This would include examining both the benefits and the burdens of making understanding explicit; clarifying its relationship to assurance and safety cases; and exploring use cases across critical systems, AI-generated assurance, more everyday systems, and different regulatory environments. It also includes extending Assurance 2.0 in areas that the present paper only sketches, especially in relation to felicitous falsehoods, novelty, distributed understanding, and practical mechanisms for demonstrating grasp.

Supporting processes will matter as much as representational structures. An understanding-centric methodology will need to be embedded in engineering workflows in ways that encourage explanation, challenge, and reflective review rather than passive acceptance of apparently coherent outputs. This is where designed friction may prove especially germane.

### A. Why this matters for governance and policy

This issue is not only technical. Assurance artefacts increasingly inform regulatory, procurement, certification, and public-sector deployment decisions. If such artefacts can be generated quickly and appear internally coherent without being well understood by those who rely on them, they may create a misleading impression of accountability, due diligence, and control. In policy settings, the core question is therefore not whether assurance documentation exists, but whether it exposes the assumptions, reasoning, uncertainties, and residual risks in a way that can be scrutinized and challenged.

For this reason, an understanding-centric approach to assurance could matter well beyond engineering teams. It could help regulators, procurers, auditors, and organizational leaders ask better questions about what is known, what is only assumed, where confidence comes from, and where judgment still depends on contestable interpretations. The proposed idea of an Understanding Basis is therefore relevant not only to technical practice but also to governance: it offers a way to make the quality of understanding more visible and to reduce the risk that automation merely industrializes persuasive paperwork.

## V. CONCLUSION

Understanding is foundational to good engineering decisions but is rarely made explicit. By integrating Elgin's epistemology with the structures of Assurance 2.0, we outline how understanding might be made explicit, systematic, defensible, and open to challenge. The *Understanding Basis* and *Personal Understanding Statement* could contribute to demonstrating and assessing understanding for complex socio-technical systems. We hope that making understanding explicit can improve transparency, challenge, robustness, and adaptability— particularly in environments where superficially coherent narratives might mask weaknesses, and where AI-supported work risks producing documentation artefacts untethered from genuine comprehension.

## VI. GLOSSARY

| Term | Engineering-Focused Definition |
| --- | --- |
| Understanding Basis | A documented body of epistemic commitments, models, and evidence that forms the foundation for a justified, defensible and revisable human grasp of system behavior. |
| Epistemic Commitment | An accountable obligation to treat a claim, model, or assumption as "true enough" for reasoning and action. |
| Grasp | The successful "take on a topic" that allows a person to wield information to reason, explain, predict, and act. |
| Felicitous Falsehood | Idealizations, models, or simplifications that are not literally true but are "true enough" to reveal salient system structures or risks. |
| Reflective Equilibrium | A state where claims, evidence, and assumptions are mutually supportive and integrated into a defensible "fabric" of understanding. |
| Exemplification | The mechanism that tethers an account to reality through samples (e.g., test results) that instantiate evidential properties. |
| Arendtian Rupture | A breakdown in existing frameworks caused by unprecedented events or radical novelty that cannot be explained by old categories of thought. |
| Rupture and re-evaluation | Recognition that established frameworks may fail in the face of radical novelty, requiring re-examination of assumptions, evidence, and judgement rather than forced coherence. |
| Epistemic Debt | A risk that functional capability, aided by automation, grows faster than the user's mental models and understanding. |
| Designed Friction | Deliberate prompts, pauses, or explanation demands introduced to support reflective review and expose shallow understanding. |